# Enhancement of the critical slowing down influenced by extended defects

V. Blavats'ka[a*], M. Dudka[a,b], R. Folk[b], and Yu. Holovatch[a,b,c]

[a] Inst. for Condensed Matter Physics, National Acad. Sci. of Ukraine, 79011 Lviv, Ukraine
[b] Institut für Theoretische Physik, Johannes Kepler Universität Linz, 4040 Linz, Austria
[c] Ivan Franko National University of Lviv, 79005 Lviv, Ukaine

We study an influence of the quenched extended defects on the critical dynamics of the $d=3$-dimensional systems with $m$-component non-conserved order parameter (model A dynamics). Considering defects to be correlated in $\varepsilon_d$ dimensions and randomly distributed in $d-\varepsilon_d$ dimensions we obtain reliable numerical values for the critical exponents governing divergence of the relaxation time as function of $m$ and $\varepsilon_d$.

Keywords: quenched disorder, extended defects, critical dynamics, renormalization

Critical slowing down constitutes one of the most prominent effects accompanying critical phenomena. Approaching the critical point $T_c$ the relaxation time $\tau$ increases and becomes infinite at $T_c$. Its divergence is related to the divergence of the correlation length $\xi$ by

$$\tau \sim \xi^z, \qquad (1)$$

where $z$ is the dynamical critical exponent $z$ [1].

Study of the influence of structural disorder on the critical behaviour is an important and non-trivial task both for the theory and the experiment. For fluids, an experimental situation may be realized by a criticality of fluids in porous medium [2]. Traditionally, such objects are modelled by $m$-vector systems with quenched uncorrelated point-like defects [3]. A more realistic description is given by models, where defects are extended in space [4,5]. Here, we present analysis of the critical dynamics for $d=3$-dimensional system with $m$-component non-conserved order parameter (model A dynamics) influenced by presence of $\varepsilon_d$-dimensional parallel defects, randomly distributed in $d$-$\varepsilon_d$ dimensions [3,4].

Reliable estimates for the exponents describing *static* critical behaviour of such a model were obtained only recently [6] exploiting results of Refs. [5,7]. Estimates of similar accuracy for the *dynamic* exponents were obtained in Ref. [8]. Here, we continue studies initiated in Ref. [8].

Presence of the extended defects of parallel orientation introduces anisotropy and two correlation lengths naturally arise in system description: one is perpendicular to the extended impurities direction, $\xi_\perp$, and the second one is parallel, $\xi_\parallel$ [3]. This modifies dynamical scaling as well. Indeed, the relaxation time in direction perpendicular to the impurities extension, $\tau_\perp$, differs from that in direction along the impurities extension, $\tau_\parallel$ and their scaling approaching the critical point is governed by:

$$\tau_\perp \sim \xi_\perp^{z_\perp}, \quad \tau_\parallel \sim \xi_\parallel^{z_\parallel} \qquad (2)$$

with two independent dynamical exponents $z_\perp$ and $z_\parallel$ [5].

To analyse asymptotic scaling behaviour (2) we make use of the standard tools of the field-theoretical renormalization group (RG) in minimal subtraction scheme [9]. We work within Bausch-Janssen-Wagner formalism [10], which reduces analysis of the dynamical equations of motion for the order parameter to the study of corresponding Lagrangian. Renormalising its parameters, such as bare couplings $u_0$, $v_0$, order parameter relaxation coefficient $\lambda_0$ and anisotropy constant $a_0$ we obtain RG functions $\beta_u(u,v)$, $\beta_v(u,v)$, $\zeta(u,v)$, $\zeta_a(u,v)$ needed to calculate dynamical critical exponents. The $\beta$-functions are characterized by presence of the stable and an accessible fixed point (FP) $u^*,v^*$ of RG transformation, which corresponds to the critical point. And in this FP critical exponents (2) are calculated from the relations:

$$z_\perp = 2 + \zeta(u^*, v^*) \tag{3}$$

$$z_\parallel = z_\perp / (1 - \zeta_a(u^*, v^*)/2) \tag{4}$$

Here we use the two-loop RG functions of Refs. [5,7]. They are obtained as expansions in powers of the renormalized couplings $u,v$. The RG perturbative expansions are known to be divergent, thus the application of resummation is required [11] to get reliable data on their basis. We apply the Chisholm-Borel resummation [9,11] to obtain dynamical critical exponents $z_\perp$ from (3). And as far as the series for $z_\parallel$ (4) contains only two terms we use for it only direct substitution of the FP coordinates.

In the Fig. 1 we present the obtained results for the critical exponents $z_\perp$ and $z_\parallel$ as functions of impurity dimension $\varepsilon_d$ for different $m$. As it is seen from the figure, with an increase of $\varepsilon_d$, for $m=2, 3, 4$ the critical exponents remain constant and equal to the corresponding exponents of pure model, until $\varepsilon_d$ is smaller than $\varepsilon_d^c(m)$. For $\varepsilon_d > \varepsilon_d^c(m)$ they begin to increase. For the Ising system ($m=1$) the values of exponents start to increase as soon as $\varepsilon_d > 0$. This behaviour is explained by generalisation of usual Harris criterion [12] for extended defects [4]. It states that the extended defects become relevant for $\varepsilon_d > d - 2/\nu_p$, where $\nu_p$ is the correlation length critical exponent of a system without defects. The above relation defines for each $\varepsilon_d$ the critical value $m_c$, below which the extended defects influences universal critical properties. It turns out, that the disorder with extended defects is relevant for $d=3$ over a wider range of $m$ than the point defect disorder.

The typical numerical values for the dynamical critical exponents of Fig. 1 are larger then those of the pure $d=3$ Ising model ($z=2.18$ [13]) as well as for the $d=3$ Ising model with point-like defects ($z=2.017$ [14]). It means that the behaviour of the relaxation time in the vicinity of a critical point is characterized by a stronger singularity. Therefore presence of extended defects leads to further enhancement of the critical slowing down in comparison with a pure system.

This work was supported by Fonds zur Förderung der wissenschaftlichen Forschung under Project Nos. P16574, 15247-PHY.

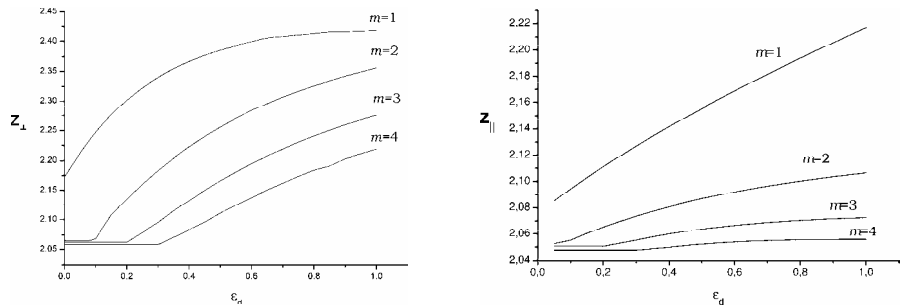

**Fig.1 The values of dynamical critical exponents $z_\perp$ and $z_\parallel$ of three-dimensional $m$-component magnets at different fixed values of extended defect dimension $\varepsilon_d$.**

*E-mail: viktoria@icmp.lviv.ua

**REFERENCES**

[1] B. I. Halperin, P. C. Hohenberg, Rev. Mod. Phys., 49 (1977) 436.
[2] A. P. Y. Wong, S. B. Kim, W. I. Goldburg, M. H. W. Chan, Phys. Rev. Lett. 70 (1993) 954; J. Yoon, D. Sergatskov, J. Ma, N. Mulders, M. H. W. Chan, Phys. Rev. Lett, 80 (1998) 1461, C. Vásquez R., P. Paredes V., A. Hasmy, R. Jullien, Phys. Rev. Lett, 90 (2003) 170602; C. Vásquez R., PhD thesis, Caracas, 2003.
[3] A. Pelissetto, E. Vicari, Phys. Rep. 368 (2002) 549.
[4] S.M. Dorogovtsev, Fiz. Tverd. Tela (Leningrad), 22 (1980) 321 [Sov. Phys.–Solid State, 22 (1980) 188]; Fiz. Tverd. Tela (Leningrad), 22 (1980) 3659 [Sov. Phys.–Solid State, 22 (1980) 2141].
[5] D. Boyanovsky and J.L. Cardy, Phys. Rev. B, 26 (1982) 154; Phys. Rev. B, 27 (1983) 6971.
[6] V. Blavats'ka, C. von Ferber, Yu. Holovatch, Acta Phys. Slovaca, 52 (2002) 317; V. Blavats'ka, C. von Ferber, Yu. Holovatch, Phys. Rev. B., 67 (2003) 094404.
[7] I. D. Lawrie and V.V. Prudnikov, J. Phys. C: Solid State Phys., 17 (1984) 1655 .
[8] V. Blavats'ka, M. Dudka, R. Folk, Yu. Holovatch, Phys. Rev. B. (2005) .
[9] see e.g. J. Zinn-Justin, *Quantum Field Theory and Critical Phenomena* (Oxford University Press, Oxford, 1996); H. Kleinert and V. Schulte-Frohlinde,*Critical Properties of $\phi^4$-Theories* (World Scientific, Singapore, 2001).
[10] R. Bausch, H. K. Janssen, and H. Wagner, Z. Phys. B, 24 (1976) 113.
[11] H. J. Hardy, *Divergent Series* (Clarendon Press, Oxford, 1949).
[12] A.B. Harris, J. Phys. C , 7 (1974) 1671.
[13] V. V. Prudnikov, A. V. Ivanov, A. A Fedorenko, Pis'ma Zh. Eksp. Teor. Fiz., 66 (1997) 793.
[14] K. Oerding and H. K. Janssen, J. Phys A, 28 (1995) 4371.